\documentclass[prb,preprintnumbers,printfigures,twocolumn,superscriptaddress,aps]{revtex4-1}
\usepackage{graphicx}
\usepackage{amsmath}
\usepackage{stackrel,amssymb,amsmath}
\usepackage{bm}
\usepackage[T1]{fontenc}
\usepackage{lmodern}
 \usepackage{textcomp}
\usepackage{amstext}
\usepackage[Symbol]{upgreek}
\usepackage{subfigure}
\usepackage{booktabs}
\usepackage{dcolumn}
\usepackage{color}
\usepackage{enumerate}
\usepackage{latexsym,bm}
\usepackage{longtable}
\usepackage[colorlinks=true, allcolors=blue]{hyperref}
\makeatletter
\begin{document}
\preprint{manuscript}
\title{Group Theory Analysis of Phonons in Monolayer Chromium Trihalides and Their Janus Structures}
\author{Y. C. Liu}
\thanks{liuyachao@xaut.edu.cn. (Y. C. Liu)}
\affiliation{Department of Applied Physics, Xi'an University of Technology, Xi'an 710054, China}

\author{H. B. Niu}
\affiliation{Department of Applied Physics, Xi'an Jiaotong University City College, Xi'an 710018, China}

\author{J. B. Lin}
\affiliation{National Institute for Materials Science, Tsukuba, 305-0044, Japan}

\author{V. Wang}
\thanks{wangvei@icloud.com. (V. Wang)}
\affiliation{Department of Applied Physics, Xi'an University of Technology, Xi'an 710054, China}

 \date{\today}
\begin{abstract}
A contrastive investigation of the symmetry aspects of phonons in monolayer chromium trihalides and their Janus structures Y$_3$-Cr$_2$-X$_3$ (X, Y = F, Cl, Br, I) by group theory is presented. We first classify all phonons at the Brillouin-zone center ($\Gamma$) into the irreducible representation. Then the infrared and Raman activity of optic phonons, Raman tensors, and the possible polarization assignments of R active phonons are predicted. Base on these results, we clarify the the discrepancy about the Raman activity o optic modes in monolayer CrI$_3$. Besides, we find that the Raman and infrared spectra for X$_3$-Cr$_2$-X$_3$ are exclusive, whereas that for Janus Y$_3$-Cr$_2$-X$_3$ are coincident. This distinction is vital for optic spectra identification of Janus Y$_3$-Cr$_2$-X$_3$ monolayer from X$_3$-Cr$_2$-X$_3$ monolayer. In addition, we derive the symmetry-matched phonon eigenfunctions and corresponding schematic representations of the eigenvectors for both F$_3$-Cr$_2$-I$_3$ and I$_3$-Cr$_2$-I$_3$ monolayer, which demonstrate intuitively the origin of phonon chirality and magnetism. At last, our analysis indicates that the spin-phonon coupling, the magneto-optical effect of infrared and Raman active phonons, and phonon chirality should be observed in Janus Y$_3$-Cr$_2$-X$_3$ monolayer as that and even easier than that in X$_3$-Cr$_2$-X$_3$ monolayer. Our work provides a detailed guiding map for experimental characterization of Y$_3$-Cr$_2$-X$_3$ monolayer, and also reveals important effects of optic phonons in Janus Y$_3$-Cr$_2$-X$_3$ monolayer.

\end{abstract}
\maketitle

\section{Introduction}
 Since the experimental demonstration of magnetism in two-dimensional (2D) CrI$_3$ monolayer,\cite{mcguire2015coupling,huang2017layer} its magnetic anisotropy and stacking-dependence, as well as strain and electric-field control, have been extensively and deeply studied in recent years.\cite{lado2017origin,sivadas2018stacking,huang2018electrical,jiang2018spin,soriano2019interplay,vishkayi2020strain-CrI3} Recently, various phonon-related effects in single- and multi-layer CrI$_3$ have also been explored theoretically and experimentally.\cite{mccreary2020distinct,huang2020tuning-via-symmetry-CrI3,yang2020magneto,jin2020observation, Magnetic-phonon-CrI3-2021}  
 
Larson and Kaxiras have obtained the frequencies and irreducible representations of the optic phonon at the Brillouin-zone center ($\Gamma$) in bulk and monolayer CrI$_3$ by an ab initio study.\cite{larson2018raman} While Webster et.al. investigated the spin-lattice and spin-phonon interactions in monolayer magnetic CrI$_3$ and obtained the phonon band dispersion as well as the schematic representation of the eigenvectors for all phonon vibrational modes at $\Gamma$ point.\cite{webster2018distinct} Some discrepancies, however, still exist in the symmetry classification of crystal vibrational modes, the identification of Raman (R), and infrared (IR) activity of optic phonons in CrI$_3$ monolayer. Thus a systematic group theory analysis of phonons in monolayer CrX$_3$ with (X = F, Cl, Br, I) is still necessary to clarify these issues. 
 
 At about the same time, the successful synthesis of Janus monolayer of transition metal dichalcogenides\cite{lu2017janus} and the prediction of their excellent electronic and piezoelectric properties\cite{dong2017large,yagmurcukardes2019electronic,chen2019symmetry,rawat2020nanoscale} have naturally stimulated further investigation of related Janus materials.\cite{li2021electronic,sun2021induced} The magnetic CrI$_3$ monolayer is no exception. The electronic structure and magnetic anisotropy of Janus CrX$_3$ monolayers have been investigated recently\cite{zhang2020spin,zhang2020electronic,albaridy2020tunable} Meanwhile, our latest work predicts a large vertical electric polarization of up to $-0.155\times10^{-10}$ C/m in Janus CrX$_3$ monolayer due to the absence of inversion symmetry, where we have provided its phonon dispersion relation.\cite{Janus-Cr2I3F3-Niu} However, so far as we know, a systematical group theory analysis of phonons in Janus CrX$_3$ monolayer is still absent in the literature. 
 
In addition, theoretical prediction of chiral phonons\cite{Zhang2015Chiral-phonon} and its soon experimental verification in tungsten-diselenide monolayers\cite{zhu2018observation} have inspired the new enthusiasm to explore chiral phonons in other 2D as well as three-dimensional (3D) materials.\cite{Chen2018-Chiral-phonons,chen2019entanglement,chen2019chiral,Liu2019ValleyselectiveCP,suri2021chiral,wang2022chiral,zhang2022chiral} For example, Yin et.al. have shown that magnetic CrBr$_3$ monolayer hosts chiral phonons at the Brillouin-zone center.\cite{Chial-phonon-CrBr3-2021} These chiral phonons are special linear combinations of the doubly-degenerate $E_g$ phonons, which eigenvectors exhibit clockwise and counterclockwise rotations with angular momentum, and can completely switch the polarization of incident circularly polarized light. Then a question naturally arises: can chiral phonons be observed in Janus CrX$_3$ with (X = F, Cl, Br, I) monolayer, and what are their unique properties that distinguish them from that in inversion-symmetric CrX$_3$ monolayer? 

In this study, the above questions would be addressed. We first perform a comprehensive and systematic group theory analysis of phonons in chromium trihalides monolayers and their Janus structures denoted by an intuitive form Y$_3$-Cr$_2$-X$_3$ with (X, Y)$\in$(F, Cl, Br, I). The irreducible representations and eigenvectors of phonons at $\Gamma$ point for inversion-asymmetric Janus Y$_3$-Cr$_2$-X$_3$ and inversion-symmetric X$_3$-Cr$_2$-X$_3$ monolayer are deduced. The IR and R activity of phonons, as well as Raman tensors of polarization setup of R active phonons, are presented. Besides, the compatibility relation of phonon irreducible representations between Y$_3$-Cr$_2$-X$_3$ and X$_3$-Cr$_2$-X$_3$ monolayer is also offered. At last, we discuss the spin-phonon coupling, the magneto-optical effect of IR and R active phonons as well as the chiral phonons in Y$_3$-Cr$_2$-X$_3$ and X$_3$-Cr$_2$-X$_3$ monolayer. Our work clarifies the R activity of optic phonons in CrI$_3$ monolayer and offers systematic spectral information about phonons in Janus Y$_3$-Cr$_2$-X$_3$ and their related effects, which are essential for theirs optical spectra identification and characterization. 

\section{Symmetry analysis of Y$_3$-C$\text{r}_2$-X$_3$ monolayer}
\subsection{Real space symmetry and wave vector group at Brillouin zone center}
Laser Raman scattering and infrared absorption spectra are both powerful tools for structural identification and characterization of 2D materials. The group theory classification of phonons in 2D materials, which has been successfully applied to graphene, transition metal dichalcogenides, and phosphorene,\cite{malard2009group,ribeiro2015group,lu2016lattice} can provide the irreducible representations of all phonons at $\Gamma$ point as well as their IR and R activity. They are essential information for the polarization setup of incident (scattering) light and the identification of optical spectra in infrared and Raman spectroscopy, which are the standard and powerful optical techniques to characterize the properties of 2D materials. 

\begin{figure}[htbp]
\centering
\includegraphics[scale=0.54]{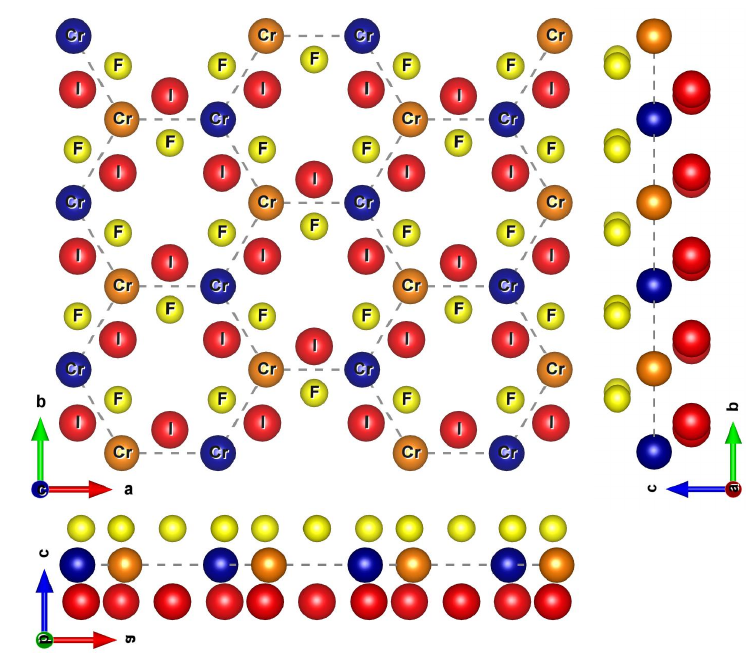}
\caption{\label{lattice structure} (Color online) The lattice structure of monolayer F$_3$-Cr$_2$-I$_3$, where green and purple spheres with the same size represent Cr atoms on different sites, while yellow and red spheres with different size represent F and I atoms, respectively, as indicated on each atom. This color convention applies to all subsequent phonon oscillations figures.}
\end{figure}

In order to guide the future experimental research, we first derive the symmetry classification of phonon modes at the $\Gamma$ point and point out IR and R activity of the optic modes in Janus Y$_3$-Cr$_2$-X$_3$ monolayer. The unit cell of Y$_3$-Cr$_2$-X$_3$ monolayer consists of three Y and three X atoms as well as two Cr atoms, in a total of eight atoms, thus there are 24 phonon modes (3 acoustic and 21 optic modes) at the $\Gamma$ point. Taking F$_3$-Cr$_2$-I$_3$ monolayer as an example, whose crystal structure is shown in Fig. \ref{lattice structure}. We have demonsted its crystal dymanical stability since there is no imaginary frequency in the phonon dispersion of the Janus F$_3$-Cr$_2$-I$_3$ monolayer.\cite{Janus-Cr2I3F3-Niu} Generally, lattice vibrational modes can be classified based on the irreducible representation of the space group.\cite{Dresselhaus2007group} The structural symmetry of Janus Y$_3$-Cr$_2$-X$_3$ monolayer belongs to the symmorphic space group No.157. It has the symmetry designation $C^{2}_{3v}$ in accord with the Schoenflies notation, and $P31m$ in the Hermann-Mauguin notation. 

The rotational aspects for real space and for the group of the wave vector at $k=0$ in reciprocal space ($\Gamma$ point) are described by the point group $C_{3v}$. Its character table is given in Table \ref{character},
\begin{table}[htbp]
\centering
 \begin{ruledtabular}
\caption{\label{character}Character table for the point group $C_{3v}$ including basis functions of the irreducible representations.}
\begin{tabular}{ccccccc}
   $\rm{SG}$&$\rm{PG}$&$E$ &$ 2C_{3}$&$3\sigma_{v}$&$\rm{Basis components}$\\
    \hline
   $\Gamma_{1}$  &$A_{1}$  &1   &{ 1}  &{ 1}    &$z, z^{2}, x^{2}+y^{2}$\\
   $\Gamma_{2}$  &$A_{2}$  &1   &{ 1}  &{-1}    &$R_{z},$\\
   $\Gamma_{3}$  &$E$        &2   &{ -1}  &{0}     &$(x,y),  (R_{x},R_{y}), (xz,yz), (x^{2}-y^{2},xy)$ \\
      \end{tabular}
 \end{ruledtabular}
\end{table}
where $\Gamma_{i}(i=1,2,3)$ are Bethe symbols of space group for irreducible representation at $\Gamma$ point, $A$ and $E$ are Mulliken signs of point group of one--dimensional (1D) and 2D irreducible representations; the subscripts $1$ and $2$ denote representations that are symmetric and antisymmetric with respect to the mirror reflection operation $\sigma_{v}$; $x$, $y$, $z$ and $R_{x}$,$R_{y}$,$R_{z}$ are basis components of polar and axial vectors, respectively. Table \ref{character} shows that $C_{3v}$ has only three irreducible representations.
\begin{table}[htbp]
\centering
\begin{ruledtabular}
\caption{\label{Reduced Representation} Characters of vector, equivalent, and vibration representations for Y$_3$-Cr$_2$-X$_3$ in $C^{2}_{3v}$ crystal structure as well as characters for X, Y and Cr atoms on different Wyckoff sites denoted by $2b$ and $3c$.}
\begin{tabular}{ccccc}
 \centering
   $C_{3v}$&$ E$ &$ C_{3}$ &$\sigma_{v}$ \\
             \hline
  $ \chi _{\rm{vector}}$&{ 3} &0 & 1 \\
  $ \chi _{\rm{equivalent}}$ &{ 8} &2 & 2 \\
   $ \chi _{\rm{vibration}}$ &24 & 0 & 2 \\
   $ \chi _{\rm{X}}(3c)$&{ 3} &0 & 1 \\
  $\chi _{\rm{Y}}(3c)$&{ 3} &0 & 1 \\
  $ \chi _{\rm{Cr}}(2b)$&{ 2} &2 & 0\\
\end{tabular}
 \end{ruledtabular}
  \end{table}
  
\subsection{Irreducible representations for vibrational modes in Y$_3$-Cr$_2$-X$_3$ monolayer}  
The lattice vibrational modes of Y$_3$-Cr$_2$-X$_3$ at $\Gamma$ are classified according to the irreducible representations of C$_{3v}$. Characters of atomic displacement vector representations, primitive cell equivalent representations, and lattice vibration representations of Y$_3$-Cr$_2$-X$_3$ have listed in Table \ref{Reduced Representation}. These representations can be reduced into the irreducible representations summarized in Table \ref{character}:
\begin{align}
\label{eq4}
\Gamma _{\rm{vector}}&= {A_1} \oplus{E},\\
\label{eq5}
\Gamma _{\rm{equivalent}}& = 3{A_1}\oplus{A_2}\oplus2{E},\\
\label{eq6}
\Gamma _{\rm{vibration}} &=\Gamma_{\rm{vector}}\otimes\Gamma_{\rm{equivalent}}\nonumber\\
                                  &=({A_1} \oplus {E})\otimes(3{A_1}\oplus{A_2}\oplus2{E})\nonumber \\
                                  &=5{A_1} \oplus 3{A_2} \oplus 8{E},\\
\Gamma _{\rm{vibration}}^{z} &={A_1}\otimes(3{A_1}\oplus{A_2}\oplus2{E})\nonumber \\
                                                &=3{A_1}\oplus{A_2}\oplus2{E},\label{eqz}\\
 \Gamma _{\rm{vibration}}^{(x,y)} &={E}\otimes(3{A_1}\oplus{A_2}\oplus2{E})\nonumber \\
                                                      &= 2{A_1}\oplus2{A_2}\oplus6{E}\label{eqxy},                                            
\end{align}
where $\Gamma _{\rm{vector}}$, $\Gamma _{\rm{equivalent}}$, and $\Gamma _{\rm{vibration}}$ are the symmetry representations of atomic displacement vector, the equivalent representations of the primitive cell and the symmetry representations of lattice vibration at $\Gamma$ point, respectively. The equivalent representation denotes the number of atoms that are invariant under the symmetry operations of the group. The symmetry representation of lattice vibration is equal to the direct product of the symmetry representations of atomic displacement vector and the equivalent representations of the primitive cell.\cite{Dresselhaus2007group,Liu2017} We show that lattice vibration at zone center in monolayer Y$_3$-Cr$_2$-X$_3$ includes 24 phonon modes, in which there are 8 non-degenerate modes(5$A_1$+3$A_2$) and 8 doubly degenerate E modes. Note that the vector transforming as $A_{1}\oplus E$ is corresponding to z and (x, y) components, it permits us to separate out the lattice modes vibrating along z-direction from those in the x-y plane. Eq. \eqref{eqz} and Eq. \eqref{eqxy} indicate the irreducible representations for out-of-plane and in-plane modes, respectively

\section{Irreducible representations for optic modes in Y$_3$-C$\text{r}_2$-X$_3$ monolayer}
The symmetry representation of lattice vibration includes 24 phonon modes entirely and can be further decomposed into the representations of acoustic and optic modes as follows:
 \begin{align}
   \label{eq7}
    \Gamma _{\rm{acoustic}}& =A_{1} \oplus E,\\
    \label{eq8}
    \Gamma _{\rm{optic}} &=\Gamma _{\rm{vibration}}-\Gamma _{\rm{acoustic}}\nonumber\\
    &=\{4{A_1}\oplus7{E}\}(\mathrm{IR}+\mathrm{R})\oplus3{A_2}(\mathrm{silent}),\\
    \label{eqoz}
    \Gamma _{\rm{optic}}^{z} &=2{A_1}\oplus{A_2}\oplus2{E}\nonumber\\
                                             &=\{2[{A_1}\oplus{E}]\}(\mathrm{X}+\mathrm{Y})\oplus{A_2}(\mathrm{Cr}),\\
         \label{eqoxy}                                       
    \Gamma _{\rm{optic}}^{(x,y)} &= 2{A_1}\oplus2{A_2}\oplus5{E}\nonumber\\
     &=\{2[{A_1}\oplus{E}]\oplus2[{A_2}\oplus{E}]\}(\mathrm{X}+\mathrm{Y})\oplus{E}(\mathrm{Cr}).\nonumber\\
     \end{align}
The acoustic modes transform as a polar vector and thus include one $A_{1}$ and one $E$ modes, all their frequencies are identical to zero. The rest of the 21 nonzero frequency modes belong to optic modes, which include four $A_{1}$ and three $A_{2}$ non-degenerate modes as well as seven doubly degenerate $E$ modes. For Cr atoms there is one out-of-plane (z) $A_2$ mode and one in-plane (x-y) E mode; for six halogen atoms (3X+3Y),
there are two pairs of $A_{1}\oplus{E} $ out-of-plane modes as well as four pairs of $A_{1(2)}\oplus{E}$ in-plane modes.

\subsection{IR and R activity of optic modes in Y$_3$-Cr$_2$-X$_3$ monolayer}   
The IR active modes are symmetry-adapted modes transforming also according to the vector representation, while the R active modes are symmetry-adapted modes transforming according to the components of a symmetric second-rank tensor representation corresponding to quadratic basis functions. Judging from the basis functions shown in Table \ref{character}, one can find that the eighteen optic modes $\{4A_{1}\oplus7{E}\}$ are both R and IR active. The left three ${A_2}$ optic modes are neither R nor IR active, which can not be detected by first-order Raman and infrared optical spectra and are thus called silent modes. However, all the three silent ${A_2}$ modes can be detected in the second-order Raman spectrum since the overtone of ${A_2}$ modes becomes a R active mode $A_1$ (${A_2}\otimes{A_2}={A_1}$). In fact, for Janus X$_3$-Cr$_2$-Y$_3$, all the fundamental optic modes at $\Gamma$ point are second-order R active modes. 

 From Eq. \eqref{eqoz} we find that for seven out-of-plane modes, there are six optic modes, two ${A_1}$ and two degenerate ${E}$ , are both infrared and Raman active which should be shown as four Raman peaks. Eq. \eqref{eqoxy} shows that there are nine in-plane modes, which also indicates that seven in-plane optic peaks are expected in both infrared and Raman spectra: two ${A_1}$ and five ${E}$. It is worthy to point out that the above symmetry analysis is suitable for all Janus phases of Y$_3$-Cr$_2$-X$_3$, with X and Y belonging to halogen elements, i.e., (F, Cl, Br, I) but without X=Y. Besides, since Y$_3$-Cr$_2$-X$_3$ monolayer is mono-axis polar material with an optic axis along the out-of-plane direction,its all R active modes, which are also IR active, become extraordinary phonons. However, the electric polarity of Y$_3$-Cr$_2$-X$_3$ does not affect the direction dispersion of the optic phonons, since its optic axis is perpendicular to the atomic monolayer and thus perpendicular to all 2D wave vectors. It is well-known that when the wave vector is parallel or perpendicular to the optic axis, the extraordinary phonons become transverse modes ( in this case $A_{1T}$ and ${E_T}$) and its electric polarization becomes zero. 
       
\subsection{Optic modes in monolayer X$_3$-Cr$_2$-X$_3$}

It is noted that the coexistence of R and IR activity for optic modes in Janus Y$_3$-Cr$_2$-X$_3$ is in sharp contrast to those of X$_3$-Cr$_2$-X$_3$, where the R and IR active modes are mutually exclusive because of the recover of inversion symmetry in the crystal. For easily identifying X$_3$-Cr$_2$-Y$_3$ structure from a Raman optical spectrum experiment, we compare the R active modes of Janus Y$_3$-Cr$_2$-X$_3$ with those of inversion-symmetric X$_3$-Cr$_2$-X$_3$, i.e., monolayer CrX$_3$. The space group of X$_3$-Cr$_2$-X$_3$ is $D^{1}_{3d}$ ($P\bar{3}1m$, No.162), whose factor group is $D_{3d}$.\cite{larson2018raman,djurdjic2018lattice} The character table for $D_{3d} $ is given in Table \ref{character-D3d},
\begin{table}[htbp]
\centering
 \begin{ruledtabular}
\caption{\label{character-D3d}Character table for the point group $D_{3d}$ including basis functions of the irreducible representations.}
\begin{tabular}{ccccccccc}    
  $\rm{SG}$ &$\rm{PG}$ &$E$  &$ 2C_{3}$ &$3C_{2}$ &$i$ &$2S_{6}$  &$3\sigma_{d}$ &$\rm{Basis}$\\
    \hline
   $\Gamma_{1}^{+}$  &$A_{1g}$  &1& {1}  &{1} &{1}& 1&1&         $z^{2},x^{2}+y^{2}$\\
   $\Gamma_{2}^{+}$  &$A_{2g}$  &1& {1}  &-1   &{1} &1&-1&           $R_{z}$\\
   $\Gamma_{3}^{+}$  &$E_{g}$    &2&  {-1}  &0   &2  &-1&0&           $(R_{x},R_{y}),(x^{2}-y^{2},xy),(xz,yz)$ \\
   $\Gamma_{1}^{-}$  &$A_{1u}$   &1&  {1}  &{1} &-1 &-1&-1 &           \\ 
   $\Gamma_{2}^{-}$  &$A_{2u}$   &1&  {1}  &-1   &{-1} &-1&1&         $z$\\
   $\Gamma_{3}^{-}$  &$E_{u}$   &2&  {-1}  &0   &-2  &1&0&             $(x,y)$ \\
   \end{tabular}
 \end{ruledtabular}
\end{table}
where the subscripts $g$ (gerade) and $u$ (ungerade) denote representations that are symmetric and antisymmetric with respect to the inversion operation; the other symbols are the same as that in Table \ref{character}. It is noted that point group $D_{3d}$ has six irreducible representations which is twice as many as that of $C_{3v}$. Three of them are inversion symmetric and the left three are inversion antisymmetric.

\begin{table}[htbp]
\centering
\begin{ruledtabular}
\caption{\label{Reduced Representation-D3d} Characters of vector, equivalent, and vibration representations for I$_3$-Cr$_2$-I$_3$ as well as characters for I and Cr atoms on different Wyckoff sites denoted by $6k$ and $2c$. }
\begin{tabular}{ccccccc}
 \centering
    $D_{3d}$ &$E$  &$ 2C_{3}$ &$3C_{2}$ &$i$ &$2S_{6}$  &$3\sigma_{d}$\\
    \hline
  $\chi _{\rm{vector}}$           &3    &0    &-1    &-3    &0    &1\\
  $\chi _{\rm{equivalent}}$    &8    &2    & 2     &0     &0    &2 \\
  $\chi _{\rm{vibration}}$       &24  &0    &-2    &0      &0   &2\\
  $\chi _{\rm{I}}(6k)$      &6    &0    & 0    &0      &0   &2 \\
  $\chi _{\rm{Cr}}(2c)$  &2    &2    & 2    &0      &0   &0 \\
\end{tabular}
 \end{ruledtabular}
  \end{table}

We classify the lattice vibrational modes of $X_3$-Cr$_2$-X$_3$ at $\Gamma$ by group theory based on the irreducible representations of $D_{3d}$. Characters of atomic displacement vector representations, primitive cell equivalent representations, and lattice vibration representations of I$_3$-Cr$_2$-I$_3$, for example, are shown in Table \ref{Reduced Representation-D3d}. These representations can be decomposed into the irreducible representations summarized in Table \ref{character-D3d}:
\begin{align}
\label{eq4-2}
\Gamma _{\rm{vector}}&= {A_{2u}} \oplus {E_{u}},\\
\label{eq5-2}
\Gamma _{\rm{equivalent}}& = 2{A_{1g}}\oplus{A_{1u}}\oplus{A_{2u}}\oplus{E_{g}}\oplus{E_{u}},\\
\label{eq6-2}
\Gamma _{\rm{vibration}} &=\Gamma_{\rm{vector}}\otimes\Gamma_{\rm{equivalent}}\nonumber\\
                                  &=({A_{2u}} \oplus {E_{u}})\otimes(2{A_{1g}}\oplus{A_{1u}}\oplus{A_{2u}}\oplus{E_{g}}\oplus{E_{u}})\nonumber \\
                                  &=2{A_{1g}}\oplus2{A_{2g}}\oplus{A_{1u}}\oplus3{A_{2u}}\oplus4{E_{g}}\oplus4{E_{u}},\\
\Gamma _{\rm{I}} &={A_{1g}}\oplus{A_{2u}}\oplus{E_{g}}\oplus{E_{u}},\\                                 
\Gamma _{\rm{Cr}}&=A_{1g}\oplus{A_{1u}} .                                  
\end{align}

This symmetry representation of lattice vibration still includes 24 phonon modes and can also  be decomposed into the representations of acoustic and optic modes as follows:
 \begin{align}
   \label{eq7-2}
    \Gamma _{\rm{acoustic}}& = {A_{2u}} \oplus{E_{u}},\\
    \label{eq8-2}
    \Gamma _{\rm{optic}} &= ({2{A_{1g}}\oplus 4{E_g}})\oplus 2{A_{2g}}\oplus{A_{1u}} \oplus[{2{A_{2u}}\oplus 3{E_u}}],\\
     \Gamma _{\rm{Cr}}^{z}&=(A_{1g}\oplus{A_{1u}})\otimes{A_{2u}}={A_{2u}}\oplus{A_{2g}}, \\  
     \Gamma _{\rm{Cr}}^{(x,y)}&=(A_{1g}\oplus{A_{1u}})\otimes {E_{u}}={E_{u}}\oplus{E_{g}},\\
      \Gamma _{\rm{I}}^{z}&=({A_{1g}}\oplus{A_{2u}}\oplus{E_{g}}\oplus{E_{u}})\otimes{A_{2u}}\nonumber\\
                                        &=A_{2u}\oplus{A_{1g}}\oplus{E_{u}}\oplus{E_{g}},\\
     \Gamma _{\rm{I}}^{(x,y)}&= ({A_{1g}}\oplus{A_{2u}}\oplus{E_{g}}\oplus{E_{u}})\otimes{E_{u}}\nonumber\\
                                       &={A_{1g}}\oplus{A_{2g}}\oplus{A_{1u}}\oplus{A_{2u}}\oplus2{E_{g}}\oplus2{E_{u}}.
                                \end{align} 
We point out that $({2{A_{1g}}\oplus 4{E_g}})$ are 10 Raman active modes, while $[{2{A_{2u}}\oplus 3{E_u}}]$ are 8 IR active modes. Our results about the irreducible representation of the fundamental optic modes in $X_3$-Cr$_2$-X$_3$ are in good agreement with that of CrI$_3$ monolayer obtained by Larson et.al.\cite{larson2018raman} The R and IR modes in $X_3$-Cr$_2$-X$_3$ are exclusive due to the presence of inversion symmetry in point group $D_{3d}$. Note that this exclusion principle does not always mean that all the $g$ modes are Raman active and the $u$ modes are IR active. The left 3 optic modes, $2{A_{2g}}\oplus{A_{1u}}$, are silent modes. 

Unfortunately, we find that the two silent $A_{2g}$ phonons of CrI$_3$ have been misidentified as R active modes,\cite{webster2018distinct} while the silent $A_{1u}$ phonon of CrI$_3$ monolayer has been misidentified as IR active mode.\cite{wang2021magnetic} Besides, based on our results, we find that the acoustic $A_{2u}$ has been misjudged as $A_{1u}$ in Webster et.al.'s results,\cite{webster2018distinct} and therefore the wrong number of the two modes in $\Gamma _{D_{3d}}$. The last but the most important, they also made the misidentification of irreducible representations for modes (j) 134.5 cm$^{-1}$ ($A_{2u}$) and (n) 264.7 cm$^{-1}$ ($A_{1u}$) in Fig. 4,\cite{webster2018distinct} the correct result should be $A_{1u}$ for (j) mode and $A_{2u}$ for (n) mode. Since this misidentification has misled the experimental identification of IR active modes of CrI$_3$ monolayer,\cite{wang2021magnetic,tomarchio2021-IR-exp} we commented the webster et al.'s results\cite{Liu2022comment} and they have made correction.\cite{Webster2022correction} We suggest that relavant mistakes in Ref [\onlinecite{wang2021magnetic}] and Ref [\onlinecite{tomarchio2021-IR-exp}] should also be corrected accordingly. 

In addition, we exhibit the out-of-plane and in-plane modes of Chromium and halogen atoms separately, which indicates that all the two $A_{1g}$ Raman active modes originate merely from the optic vibration of halogen atoms, while Cr atoms only take part in the Raman active $E_g$ modes. Since magnetism of CrI$_3$ comes from Cr atoms,\cite{wang2011electronic,mcguire2015coupling,huang2017layer} the distinct spin-Raman phonon coupling should be significant in the $E_g$ modes involving vibration of Cr atoms. This has been verified in linear polarized Raman spectra for monolayer CrI$_3$.\cite{larson2018raman,webster2018distinct}

\subsection{Compatibility relation between monolayer Y$_3$-Cr$_2$-X$_3$ and X$_3$-Cr$_2$-X$_3$} 
Compatibility relation for group-subgroup pairs from $D_{3d}$ to $C_{3v}$ is shown in Table \ref{Correlation relation}. Note that inversion symmetry is disappeared when the symmetry is degraded from $D_{3d}$ to $C_{3v}$ and thus the $g$ and $u$ subscripts also disappear for the notation of irreducible representation. The correlation is not trivial noting that the $A_{1u}$ becomes $A_{2}$, while $A_{2u}$ turns into $A_{1}$ since the subscript 1 and 2 in $D_{3d}$ and $C_{3v}$ have a different meaning for one-dimensional inversion antisymmetric irreducible representations. From Table \ref{character} and Table \ref{character-D3d}, one can find that the subscript 1(2) of the irreducible representation in $C_{3v}$ denotes symmetric (anti-symmetric) relative to mirror $\sigma$, while in $D_{3d}$ it denotes symmetric (anti-symmetric) relative to rotation $C_2$. For inversion symmetric representation in $D_{3d}$, subscript 1 and 2 are of the same meaning as to mirror $\sigma_d$; while for inversion antisymmetric representation, they are of the inverse meaning as to mirror $\sigma_d$. 

Using this transformation as shown in Table \ref{Correlation relation}, one can obtain the irreducible representations of vibrational modes of Y$_3$-Cr$_2$-X$_3$ with $C_{3v}$ symmetry from that of X$_3$-Cr$_2$-X$_3$ with $D_{3d}$ symmetry directly. It is found that the Raman modes $A_{1g}$ and $E_{g}$ in $D_{3d}$ transform to $A_{1}$ and $E$ in $C_{3v}$; comparatively, the IR modes $A_{2u}$ and $E_{u}$ in $D_{3d}$ also transform to $A_{1}$ and $E$. $A_{1}$ and $E$, however, are both R and IR active modes in $C_{3v}$. Therefore, we conclude that the Raman and infrared spectra for X$_3$-Cr$_2$-X$_3$ are exclusive, whereas the Raman and infrared spectra for Janus Y$_3$-Cr$_2$-X$_3$ are coincident. This distinct difference can be used to identify the Janus structure from its mother high-symmetry structure. In addition, the three silent modes $2{A_{2g}}\oplus{A_{1u}}$ in X$_3$-Cr$_2$-X$_3$ are all transformed to $A_{2}$ modes in Y$_3$-Cr$_2$-X$_3$, which still remain silent. 

\begin{table}[htbp]
\centering
 \begin{ruledtabular}
\caption{\label{Correlation relation}Correlation table for $D_{3d}$ and $C_{3v}$ point groups}
 \begin{tabular}{cccc}
 \multicolumn{2}{c}{$D_{3d}$}&\multicolumn{2}{c}{$C_{3v}$} \\ 
    \hline
   $\Gamma_{1}^{+}$  &$A_{1g} $&$\Gamma_{1}$  &$A_{1}$   \\
   $\Gamma_{2}^{+}$  &$A_{2g}    $    &$\Gamma_{2}$  &$A_{2}$   \\
   $\Gamma_{3}^{+}$  &$E_{g}      $    &$\Gamma_{3}$  &$E$       \\ 
   $\Gamma_{1}^{-}$  &$A_{1u}     $    &$\Gamma_{2}$  &$A_{2}$\\ 
   $\Gamma_{2}^{-}$  &$A_{2u}     $    &$\Gamma_{1}$  &$A_{1}$\\
   $\Gamma_{3}^{-}$  &$E_{u}       $    &$\Gamma_{3}$  &$E$\\
   \end{tabular}
 \end{ruledtabular}
\end{table} 

\subsection{Raman tensor of optic modes in monolayer Y$_3$-Cr$_2$-X$_3$ and X$_3$-Cr$_2$-X$_3$}

The use of linear polarized light plays a major role in the assignment of experimentally observed Raman peaks to specific R modes. In Raman experiments with linear polarized light, it is customary to use Porto's notation: $\mathbf{k}_{i}(\mathbf{e}_{i}\mathbf{e}_{s})\mathbf{k}_{s}$ to denote the incident propagation direction $\mathbf{k}_{i}$, the incident and scattered polarization directions $(\mathbf{e}_{i}\mathbf{e}_{s})$ and the scattered propagation direction $\mathbf{k}_{s}$, where $\mathbf{ e}_{i}$ and $\mathbf{e}_{s}$ are, respectively, the incident and the scattered electric fields. It is customary to designate the scattered light as having diagonal Raman components ($\mathbf{e}_{i}\|\mathbf{e}_{s}$), or off-diagonal Raman components $(\mathbf{e}_{i}\bot\mathbf{e}_{s})$. 

\begin{table}[htbp]
\centering
 \begin{ruledtabular}
\caption{\label{Raman tensor-1}Irreducible representations of R active modes and Raman tensor for both $C_{3v}$ and $D_{3d}$ point groups}
 \begin{tabular}{ccc}
     $A_{1}$       &$E(y)$           &$E(-x)$\\
    $A_{1g}$      &$E_{g}$      &$E_{g}$ \\ 
    $\left(\begin{array}{ccc}a &0&0 \\0&a &0\\0&0&b\end{array}\right)$&$\left(\begin{array}{ccc}c &0 &0\\0 &-c &d \\0&d &0\end{array}\right)$&$\left(\begin{array}{ccc} 0&-c &-d\\-c  &0 & 0 \\-d & 0 &0 \end{array}\right)$\\
        \hline
    $(\mathbf{e}_{i}\|\mathbf{e}_{s})$.  &$(\mathbf{e}_{i}\|\mathbf{e}_{s})$ and $(\mathbf{e}_{i}\bot\mathbf{e}_{s})$.   &$(\mathbf{e}_{i}\bot\mathbf{e}_{s})$\\
    \hline
    $(\textrm{e}_{ix},\textrm{e}_{sx})$&$(\textrm{e}_{ix},\textrm{e}_{sx})$&$(\textrm{e}_{ix},\textrm{e}_{sy})$\\
      $(\textrm{e}_{iy},\textrm{e}_{sy})$&$(\textrm{e}_{iy},\textrm{e}_{sy})$&$(\textrm{e}_{iy},\textrm{e}_{sx})$\\
        $(\textrm{e}_{iz},\textrm{e}_{sz})$&$(\textrm{e}_{iy},\textrm{e}_{sz})$&$(\textrm{e}_{ix},\textrm{e}_{sz})$\\
              $$&$(\textrm{e}_{iz},\textrm{e}_{sy})$&$(\textrm{e}_{iz},\textrm{e}_{sx})$\\
   \end{tabular}
 \end{ruledtabular}
\end{table} 

Table \ref{Raman tensor-1} shows the symmetric form of the Raman polarizability tensor and the possible polarization assignments of the incident and the scattered electric field for observing the Raman active modes with $C_{3v}$ and $D_{3d}$ point groups structures. 

 For X$_3$-Cr$_2$-X$_3$ monolayer of $D_{3d}$ point groups, there are two types of R active modes $A_{1g}$ and $E_g$. Considering both the permitted polarization assignments inTable \ref{Raman tensor-1} of the incident and the scattered electric field and Eq. \eqref{eq8-2}, we conclude that one can observe six Raman peaks (both 2$A_{1g}$ and 4$E_g$) in parallel polarization laser set-up and only four $E_g$ in perpendicular (or crossing) polarization laser set-ups. This conclusion agrees well with the other theoretical and experimental polarized Raman spectra for monolayer CrI$_3$.\cite{larson2018raman,djurdjic2018lattice,webster2018distinct}

 For Y$_3$-Cr$_2$-X$_3$ monolayer of $C_{3v}$ point groups, there are also two R active modes $A_{1}$ and $E$. Considering both the permitted polarization assignments on the bottom of Table \ref{Raman tensor-1} of the incident and the scattered electric field and Eq. \eqref{eq8-2}, we conclude that one can observe in principle eleven Raman peaks (both 4$A_{1}$ and 7$E$) in parallel polarization laser set-up and seven Raman peaks corresponding to $E$ modes in perpendicular polarization laser set-ups. Specifically, if one detects the scattered polarized light in the same polarization direction as the incident polarized light, he or she can observe both $A_1$ and $E$ modes; in contrast, if one probes the scattered polarized light in the polarization direction perpendicular to that of incident polarized light, then only degenerate $E$ modes can be observed.
   
Moreover, we note that for the R active mode with $A_1$ symmetry in the first column of Table \ref{Raman tensor-1}, the induced dipole has the same laser polarization direction as the incident electric field, thus only can be detected by the parallel set-up of polarization. For two-dimensional materials, the incident light and detector are usually installed along the $z$ direction with a back-scattering structure, and the permitted polarization is in $x$ and $y$ directions. Thus there are twelve in-plane R modes for Y$_3$-Cr$_2$-X$_3$ monolayer that may be observed by the parallel set-up of polarization, corresponding to seven Raman peaks in spectroscopy: two non-degenerate $A_1$ peaks and five doubly degenerate $E$ peaks. 

\section{Vibrational eigenvectors of optic modes}
\subsection{Eigenvectors of optic modes in F$_3$-Cr$_2$-I$_3$}
\begin{figure}[htbp]
\centering
\includegraphics[scale=0.4]{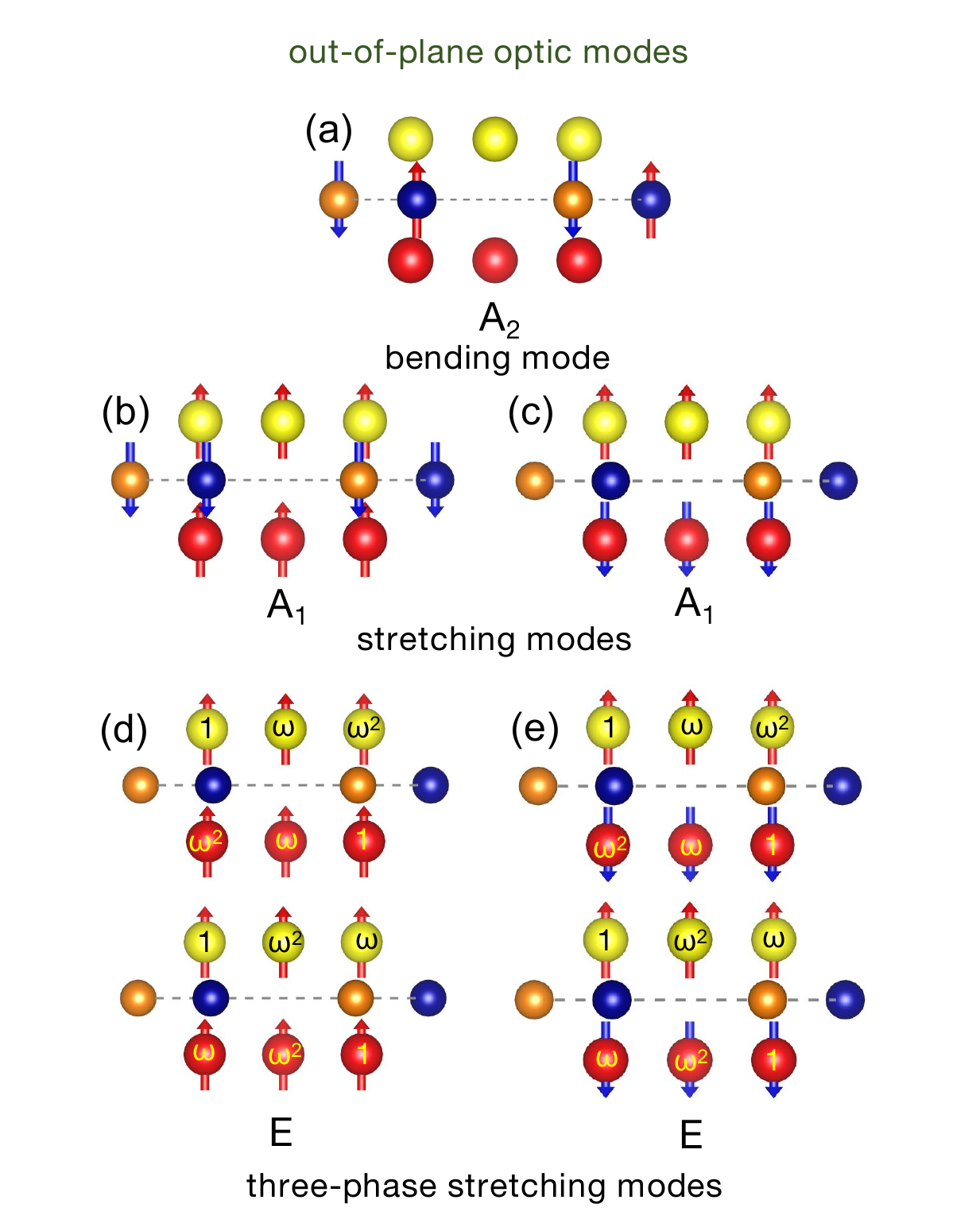}
\caption{\label{z-optical modes} (Color online) The eigenvectors of z-axis optic modes in F$_3$-Cr$_2$-I$_3$, where $\omega=e^{i2\pi/3}$, ($1,\omega, \omega^2$) indicates the relative phases of corresponding atoms; the phases of all unlabeled vibrating atoms defaults to 1; the arrows represent the direction of the atomic vibration, while their colors have no particular meaning.}
\end{figure}

The vibration directions of IR and R active modes are vital for setting the incident and detection directions as well as the polarization of the light used in optical spectra experiments. Therefore, we analyze the vibrational eigenvectors of optic modes in monolayer F$_3$-Cr$_2$-I$_3$, which is also suitable for other Janus Y$_3$-Cr$_2$-X$_3$ monolayer with same symmetry. First, the out-of-plane optic modes for monolayer F$_3$-Cr$_2$-I$_3$ are shown in Fig. \ref{z-optical modes}. One can see that the $A_2$ mode is a z-axis bending mode for two Cr atoms in the unit cell. Both the two $A_1$ modes are stretching modes along the z-direction: the one is only for halogen atoms and the other is combined in anti-phase with Cr atoms. One partner of $E$ modes is related to the corresponding $A_1$ modes by combining the $1,\omega, \omega^2$ phases with the three Fluorine and three Iodine atoms, respectively; the other partner of $E$ mode is the complex conjugate of the former and is acquired by the interchange of $\omega$ and $\omega^2$.

\begin{figure}[htbp]
\centering
\includegraphics[scale=0.42]{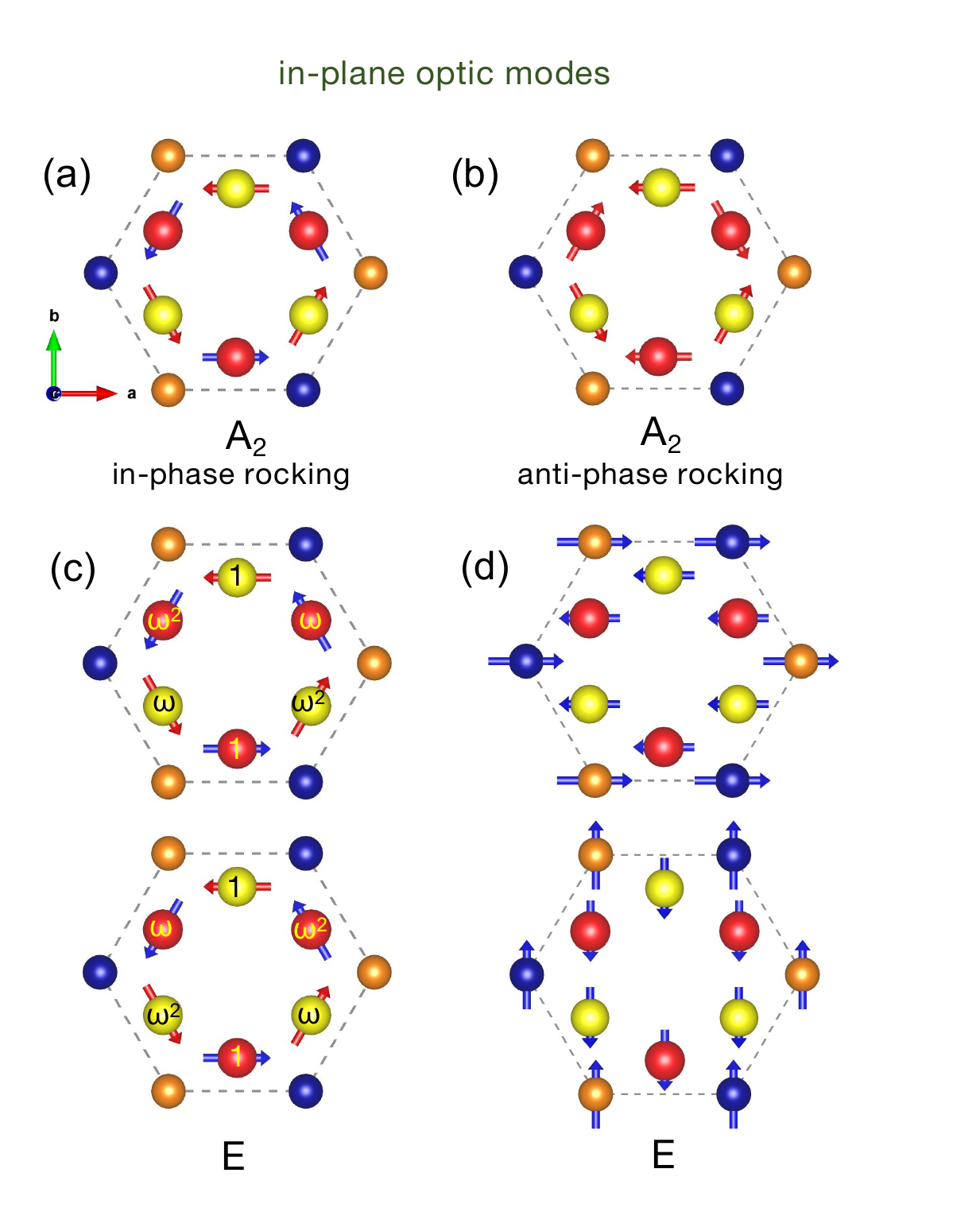}
\caption{\label{in-plane modes-1} (Color online) The eigenvectors of in-plane tangential optic modes in F$_3$-Cr$_2$-I$_3$, where $\omega=e^{i2\pi/3}$; ($1, \omega, \omega^2$) indicates the relative phases of corresponding atoms; the phases of all unlabeled vibrating atoms defaults to 1.}
\end{figure}

\begin{figure}[htbp]
\centering
\includegraphics[scale=0.42]{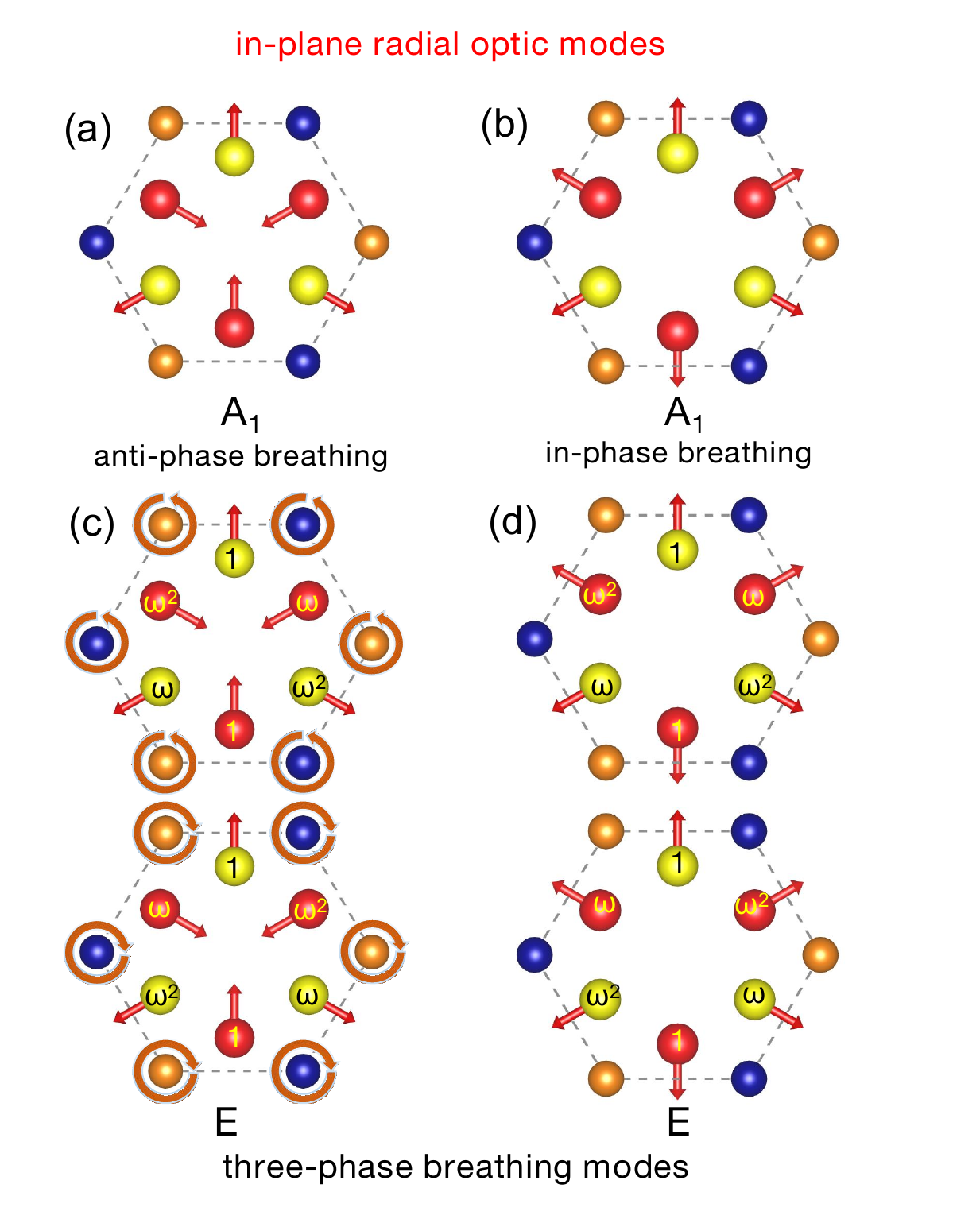}
\caption{\label{in-plane modes-2} (Color online) The eigenvectors of in-plane radial optic modes in F$_3$-Cr$_2$-I$_3$, where $\omega=e^{i2\pi/3}$; ($1, \omega, \omega^2$) indicates the relative phases of corresponding fluorine and iodine atoms; the relative phases of all unlabeled vibrating atoms defaults to 1. The circles with arrow in (c) indicate the circular movements of Cr atoms, which are necessary to keep the mass center of unit cell still.}
\end{figure}

The in-plane optic modes for halogen atoms (three Fluorine and three Iodine) in monolayer F$_3$-Cr$_2$-I$_3$ are shown in Fig. \ref{in-plane modes-1} and \ref{in-plane modes-2}. Fig. \ref{in-plane modes-1} shows the four in-plane tangential modes, including two A$_2$ and two E modes. We find that the two $A_2$ are rocking modes with one in-phase and the other anit-phase for F and I atoms. The in-phase $A_2$ mode has a corresponding E mode shown below, which is obtained similarly to that for z-axis $A_1$ in Fig. \ref{z-optical modes}(c). 

Fig. \ref{in-plane modes-2} shows the four in-plane radial modes, including two $A_1$ and two $E$ modes. It is noted that both the two $A_1$ modes are breathing modes with one in-phase and the other anit-phase for F and I atoms. The corresponding $E$ modes shown below are also obtained in the similar way as that for in-plane $A_2$ mode in Fig. \ref{in-plane modes-1}(c). Since the relative phase factors for the three F and I atoms in these $E$ modes are similar to that of three-phase alternating current, we call them 'three-phase' breathing modes. But strictly speaking, the $E$ modes stemming from the $A_1$ breathing mode in  are acturally not breathing modes because the corresponding vibrations change the symmetry of the unit cell. Besides, the circular motion of Chromium atoms in Fig. \ref{in-plane modes-2}(c) is necessary for this $E$ mode to keep the mass center from moving. It is worth pointing out that the circular motion is one origin of phonon chirality. 

\begin{figure}[htbp]
\centering
\includegraphics[scale=0.48]{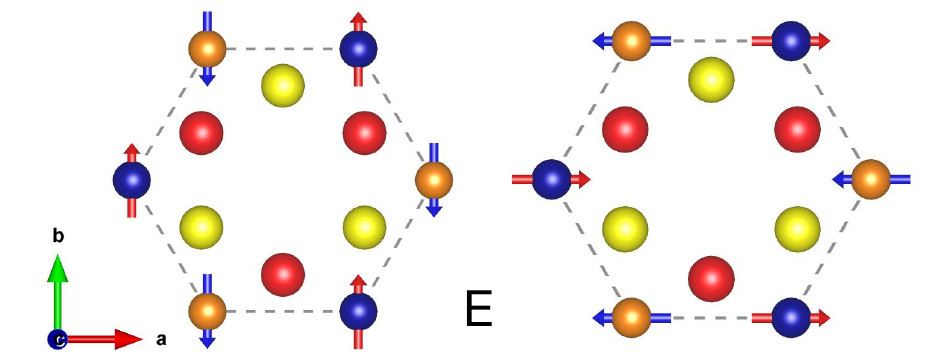}
\caption{\label{in-plane modes-3} (Color online) The eigenvectors for in-plane optic modes of Cr atoms in F$_3$-Cr$_2$-I$_3$.}
\end{figure}

If one only considers the vibration of halogen atoms in the unit cell, each $A$ mode including the motion of three F and three I atoms should have a corresponding doubly degenerate $E$ mode, constituting the three eigenfunctions of the $\hat{C}_{3}$ rotation operator by the three eigenvalues 1, $\omega$ and $\omega^2$. Since $\omega^*=\omega^2$, the two eigenfunctions with eigenvalues $\omega$ and $\omega^2$ form a doubly degenerate energy level. However, the anti-phase coupling between the acoustic vibrational modes of Cr and halogen atoms can break this correspondence. For example, the $A_1$ optic mode shown in Fig. \ref{z-optical modes}(b) and the $E$ mode shown in Fig. \ref{in-plane modes-1}(d). The in-plane $E$ mode of two Cr atoms shown in Fig. \ref{in-plane modes-3}(d) are not the eigenfunctions of $\hat{C}_{3}$ operator and thus does not have a corresponding $A$ mode ($A_1$ or $A_2$). This $E$ mode is the same as that $ E$ mode in graphene of two carbon atoms. The $A_1$ mode should be observed in both Raman and infrared spectra; the $A_2$ mode is both R and IR inactive and thus can not be detected in optical spectra. All the $E$ modes are both R and IR active and thus should be observed in optical spectra. It is pointed out that all modes with the same symmetry would be coupled to a certain degree so that any actual mode should be an admixture of the modes with the same irreducible representation. The specific value of vibrational frequency and direction for mixed mode can not be obtained by symmetry analysis.

\subsection{Eigenvectors of optic modes in I$_3$-Cr$_2$-I$_3$}
\begin{figure}[htbp]
\centering
\includegraphics[scale=0.42]{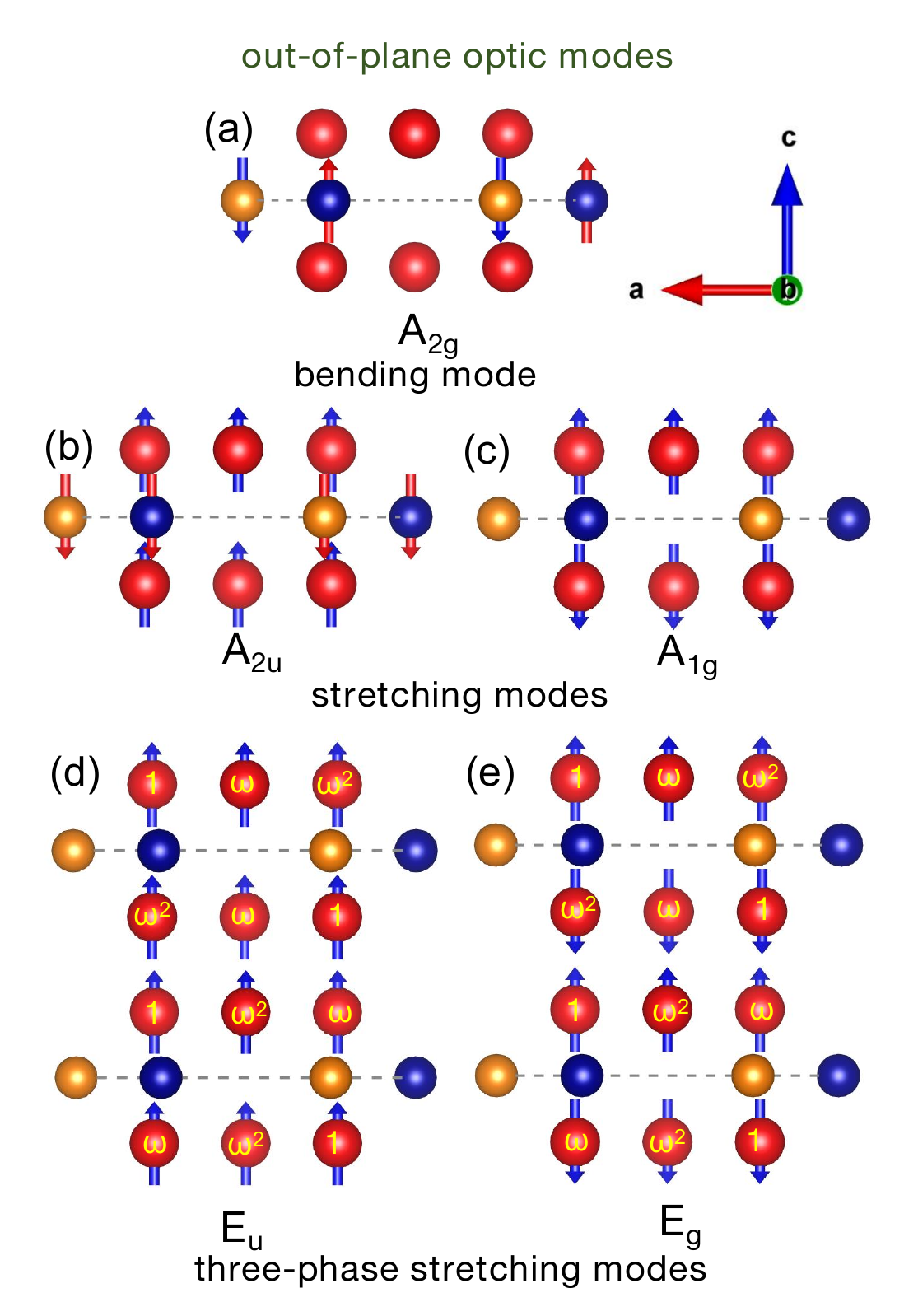}
\caption{\label{z-optical modes-b} (Color online) The eigenvectors of z-axis optic modes in I$_3$-Cr$_2$-I$_3$. Here $\omega=e^{i2\pi/3}$, ($1,\omega, \omega^2$) indicate the relative phases of corresponding atoms; the phases of all unlabeled vibrating atoms defaults to 1.}
\end{figure}

\begin{figure}[htbp]
\centering
\includegraphics[scale=0.42]{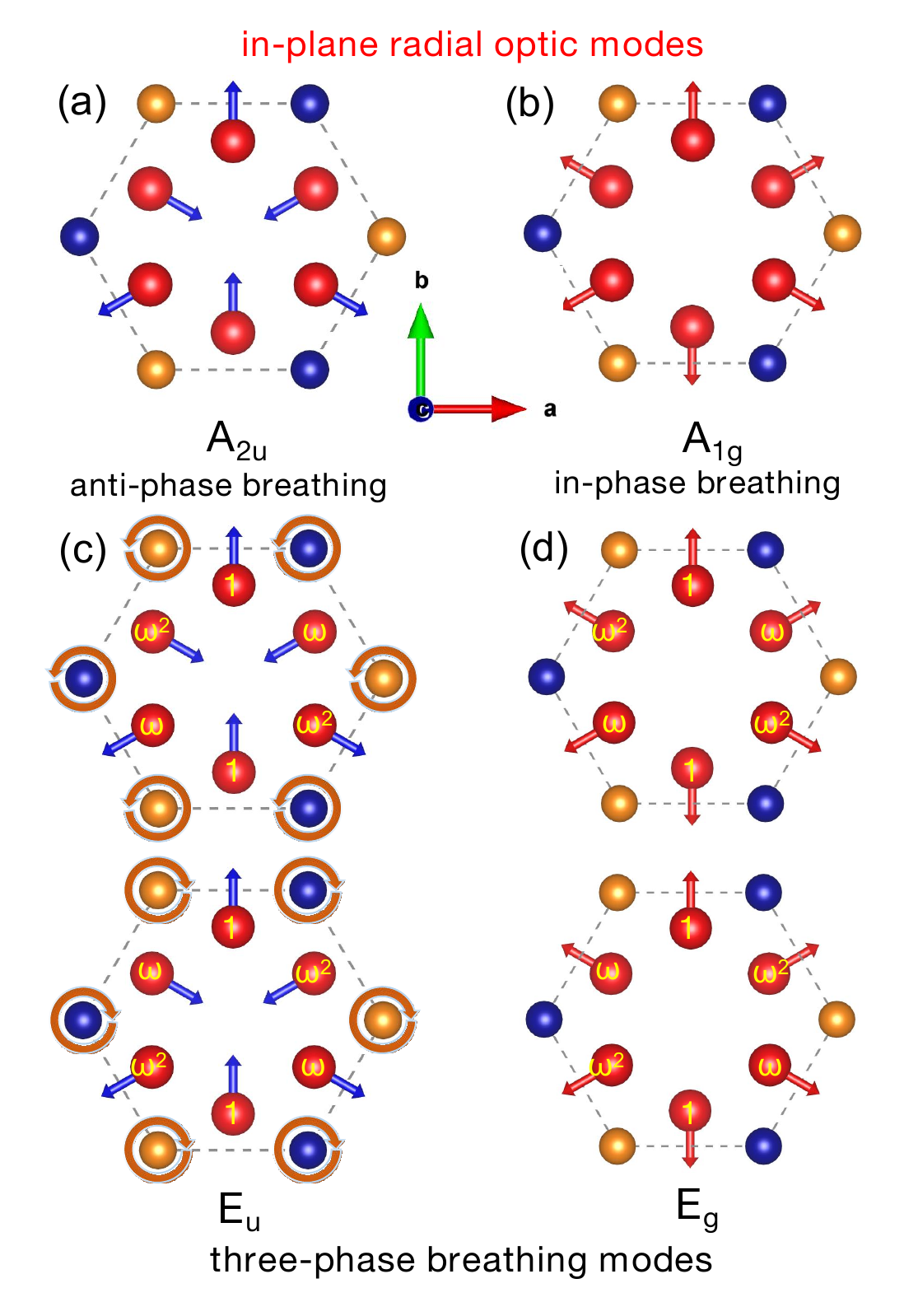}
\caption{\label{in-plane modes-2b} (Color online) The eigenvectors of in-plane radial optic modes of F and I atoms in I$_3$-Cr$_2$-I$_3$. Here $\omega=e^{i2\pi/3}$, $1,\omega, \omega^2$ indicate the relative phases of corresponding atoms; the phase of all unlabeled vibrating atoms defaults to 1. The circles with arrow in (c) indicate the circular movements of Cr atoms, which are necessary to keep the mass center of unit cell still.}
\end{figure}

\begin{figure}[htbp]
\centering
\includegraphics[scale=0.42]{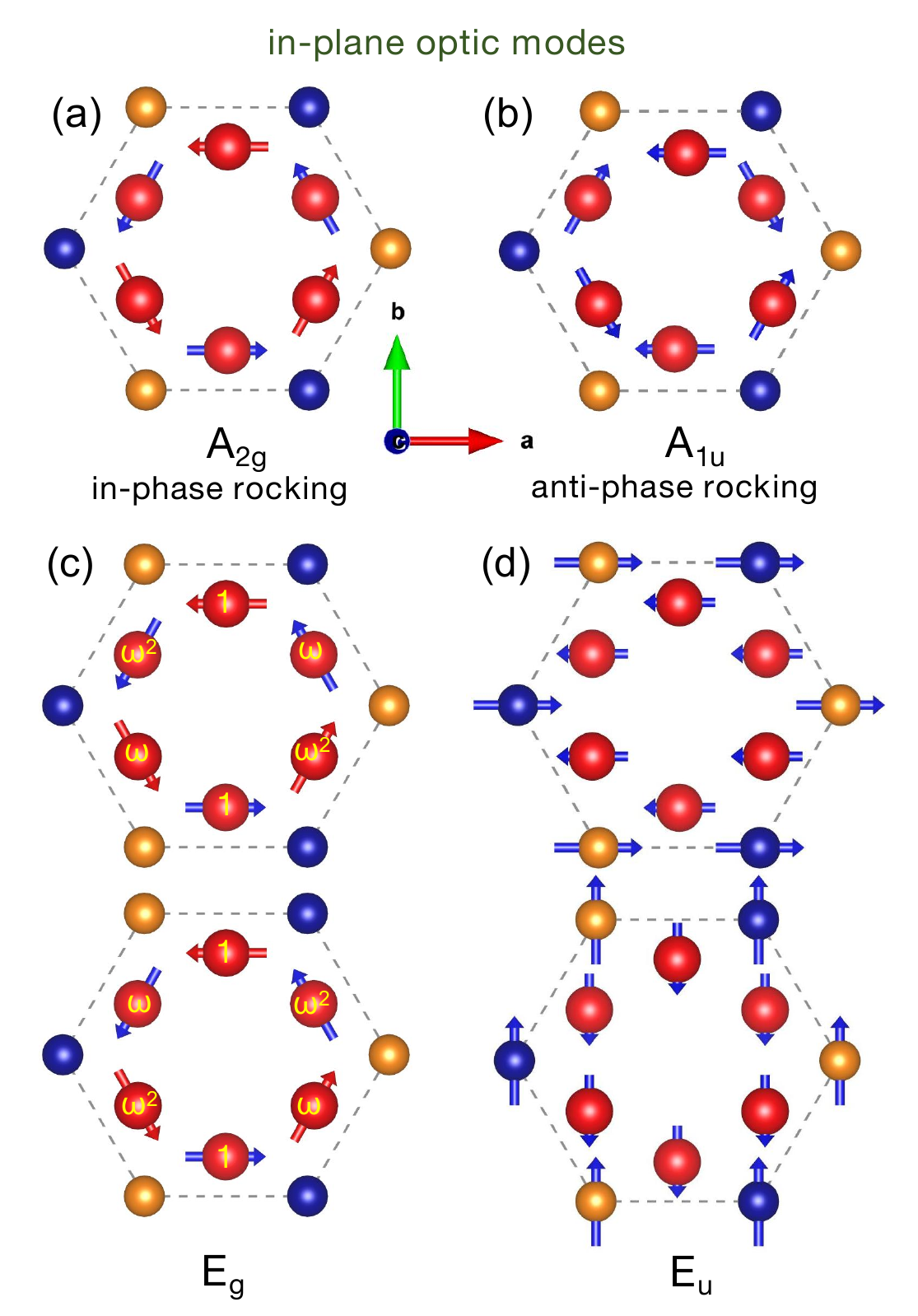}
\caption{\label{in-plane modes-1b} (Color online) The eigenvectors of in-plane tangential optic modes of F and I atoms in I$_3$-Cr$_2$-I$_3$. Here $\omega=e^{i2\pi/3}$, ($1,\omega, \omega^2$) indicates the relative phases of corresponding atoms; the phase of all unlabeled vibrating atoms defaults to 1.}
\end{figure}

Again, for the sake of comparison, we present the vibrational eigenvector of optic phonons in monolayer I$_3$-Cr$_2$-I$_3$, i.e., CrI$_3$. First, we present the out-of-plane optic modes in Fig. \ref{z-optical modes-b}. One can see that the $A_{2g}$ mode of Cr atoms of I$_3$-Cr$_2$-I$_3$ monolayer shown in Fig. \ref{z-optical modes-b}(a) is the same as the $A_2$ in F$_3$-Cr$_2$-I$_3$ shown in Fig. \ref{z-optical modes}(a) except for three F atoms substituting by three I atoms. Actually, the other two non-degenerate modes $A_{2u}$ and $A_{1g}$ in Fig. \ref{z-optical modes-b}(b,c) are also corresponding the two $A_1$ modes in Fig. \ref{z-optical modes}(b,c) only by substituting three F atoms. The eigenfunctions of non-degenerate z-axis modes in Fig. \ref{z-optical modes-b}(a-c) can be written as
\begin{align}
&A_{2g}: (z_{1}-z_{2})_{Cr},\nonumber\\
&A_{2u}: (z_{1}+z_{2}+z_{3})^{I}+(z_{1}+z_{2}+z_{3})_{I}-(z_{1}+z_{2})_{Cr}, \nonumber\\
&A_{1g}: (z_{1}+z_{2}+z_{3})^{I}-(z_{1}+z_{2}+z_{3})_{I},\nonumber\\
\end{align}
here the number subscripts represent different atoms; the element superscript and subscript denote iodine atoms on the top and bottom layers of I$_3$-Cr$_2$-I$_3$,  as well as chromium in the middle layer. The corresponding two eigenfunctions for A$_1$ out-of-plane modes of F$_3$-Cr$_2$-I$_3$ shown in Fig \ref{z-optical modes}(b,c) are simply
\begin{align}
&A_{1}(1): (z_{1}+z_{2}+z_{3})^{F}+(z_{1}+z_{2}+z_{3})_{I}-(z_{1}+z_{2})_{Cr}\label{eqA1a-z}, \nonumber\\
&A_{1}(2): (z_{1}+z_{2}+z_{3})^{F}-(z_{1}+z_{2}+z_{3})_{I},\nonumber\\
\end{align}
just with the I atoms on the top layer substituting by the F atoms. The $E_u$ and $E_g$ are simply of the same correspondence. If we write the eigenfunctions of $E_u$ and $E_g$ modes shown in Fig. \ref{z-optical modes-b}(d,e) as
\begin{align}
&E_u:\{(z_{1}+{\omega}z_{2}+\omega^{2}z_{3})^{I}+(z_{1}+{\omega}z_{2}+\omega^{2}z_{3})_{I}, c.c.\},\nonumber\\
&E_g:\{(z_{1}+{\omega}z_{2}+\omega^{2}z_{3})^{I}-(z_{1}+{\omega}z_{2}+\omega^{2}z_{3})_{I}, c.c.\},\nonumber\\
\end{align}
here c.c. means complex conjugate, then the eigenfunctions of two out-of-plane $E$ modes in Fig. \ref{z-optical modes}(d,e) should be written as 
\begin{align}
&E(1):\{(z_{1}+{\omega}z_{2}+\omega^{2}z_{3})^{F}+(z_{1}+{\omega}z_{2}+\omega^{2}z_{3})_{I}, c.c.\},\nonumber\\
&E(2):\{(z_{1}+{\omega}z_{2}+\omega^{2}z_{3})^{F}-(z_{1}+{\omega}z_{2}+\omega^{2}z_{3})_{I}, c.c.\}.
\end{align}
Actually, this correspondence is also applicable for the in-plane radial (r) three-phase breathing modes except for circular motion of Cr atoms to limit the translational motion of the center of unit cell mass. For the in-plane radial (r) optic modes in Fig. \ref{in-plane modes-2b}(a-d), the eigenfunctions of them are
\begin{align}
&A_{2u}: (r_{1}+r_{2}+r_{3})^{I}-(r_{1}+r_{2}+r_{3})_{I}\nonumber\\
&E_u:\{(r_{1}+{\omega}r_{2}+\omega^{2}r_{3})^{I}-(r_{1}+{\omega}r_{2}+\omega^{2}r_{3})_{I}+\nonumber\\
& \qquad\quad (r^{-}_{1}+r^{-}_{2})_{Cr}, c.c.\},  (r^{\pm}=x\pm iy),\nonumber\\
&A_{1g}:(r_{1}+r_{2}+r_{3})^{I}+(r_{1}+r_{2}+r_{3})_{I},\nonumber\\
&E_{g}:\{(r_{1}+{\omega}r_{2}+\omega^{2}r_{3})^{I}+(r_{1}+{\omega}r_{2}+\omega^{2}r_{3})_{I}, c.c.\}.\nonumber\\
\end{align}
Substituting the I on superscripts by F in above eigenfunctions, one would get the eigenfunctions for optic modes shown in Fig. \ref{in-plane modes-2}(a-d):\begin{align}
&A_{1}: (r_{1}+r_{2}+r_{3})^{F}-(r_{1}+r_{2}+r_{3})_{I}\nonumber\\
&E:\{(r_{1}+{\omega}r_{2}+\omega^{2}r_{3})^{F}-(r_{1}+{\omega}r_{2}+\omega^{2}r_{3})_{I}+\nonumber\\
& \qquad (r^{-}_{1}+r^{-}_{2})_{Cr}, c.c.\},  (r^{\pm}=x\pm iy),\nonumber\\
&A_{1}:(r_{1}+r_{2}+r_{3})^{F}+(r_{1}+r_{2}+r_{3})_{I},\nonumber\\
&E:\{(r_{1}+{\omega}r_{2}+\omega^{2}r_{3})^{F}+(r_{1}+{\omega}r_{2}+\omega^{2}r_{3})_{I}, c.c.\}.\nonumber\\
\end{align}
For the in-plane tangential ($t$) and rectangular coordinate axes ($x,y$) directions optic modes in Fig. \ref{in-plane modes-1b}(a-d), theirs eigenfunctions are
\begin{align}
&A_{2g}: (t_{1}+t_{2}+t_{3})^{I}+(t_{1}+t_{2}+t_{3})_{I},\nonumber\\
&E_g:\{(t_{1}+{\omega}t_{2}+\omega^{2}t_{3})^{I}+(t_{1}+{\omega}t_{2}+\omega^{2}t_{3})_{I}, c.c.\},\nonumber\\
&A_{1u}:(t_{1}+t_{2}+t_{3})^{I}-(t_{1}+t_{2}+t_{3})_{I},\nonumber\\
&E^{x}_{u}:\{(x_{1}+x_{2})_{Cr}-(x_{1}+x_{2}+x_{3})^{I}-(x_{1}+x_{2}+x_{3})_{I}\},\nonumber\\
&E^{y}_{u}:\{(y_{1}+y_{2})_{Cr}-(y_{1}+y_{2}+y_{3})^{I}-(y_{1}+y_{2}+y_{3})_{I}\}.\nonumber\\
\end{align}
The linear combinations $E^{x}_{u} \pm iE^{y}_{u}$ become the eigenfunctions of $\hat{C}_3$ operator with eigenvalues $\omega^{*}$ and $\omega$, which represent the chiral phonons with pseudo angular momentum $\mp1$.

\begin{figure}[htbp]
\centering
\includegraphics[scale=0.40]{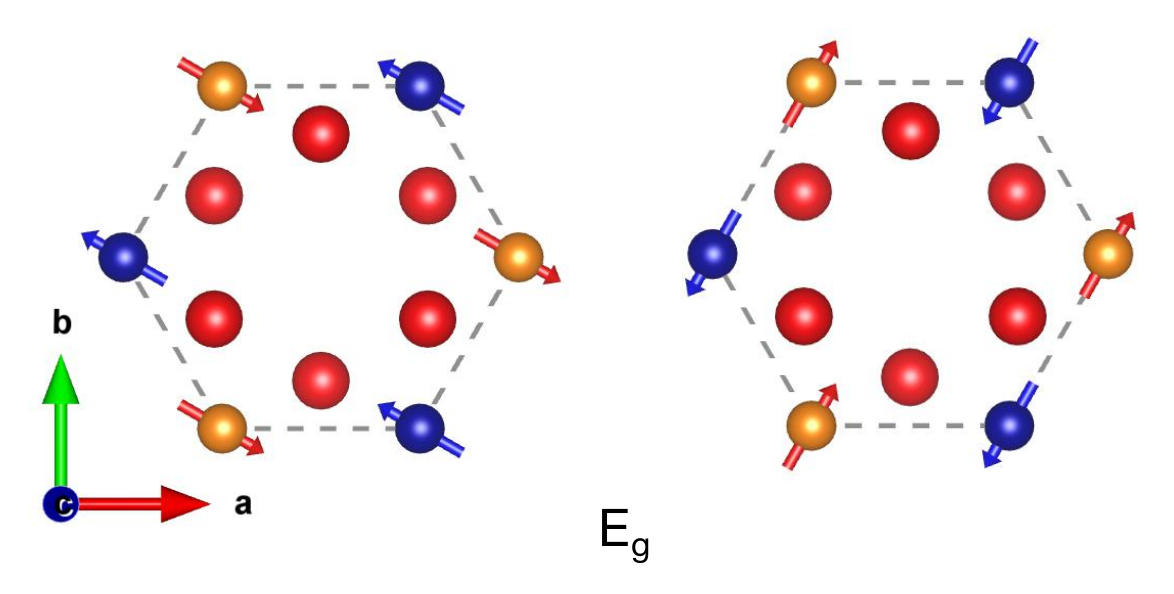}
\caption{\label{in-plane modes-3b} (Color online) The eigenvectors for in-plane optic modes of Cr atoms in I$_3$-Cr$_2$-I$_3$.}
\end{figure}

Indeed, the in-plane optic modes of Cr atoms would mix with that of halogen atoms with the same irreducible representation to some extent. Fig. \ref{in-plane modes-3b} shows the $E_g$ optic mode. It can be seen that the distance between Cr atoms varying during vibration is particularly sensitive to the magnetic order of CrI$_3$.\cite{webster2018distinct} Webster et.al. found that when the magnetic order transitions from anti-ferromagnetic (AFM) to FM, the frequency of $E_g$ modes increase while their intensity decreases, which corresponds to distinct spin-phonon interaction.\cite{webster2018distinct} The frequency increase is due to the enhancement of the effective force constant by super-exchange interaction in the ferromagnetic phase. This spin-phonon interaction is also present at the transition of interlayer magnetic order which has been reported in magneto-Raman spectroscopy study of multilayered CrI$_3$.\cite{mccreary2020distinct} 
 
\section{Discussions}

\subsection{Magnetic-phonon coupling effect of IR active phonons} 
The IR active optic phonons could couple with photons directly and thus can be detected by infrared absorption spectrometry. For X$_3$-Cr$_2$-X$_3$ ( such as CrI$_3$) monolayer, there are five IR active phonons including two non-degenerate $A_{2u}$ modes and three double-degenerate $E_{u}$ modes. As pointed out in the above section, they are Raman inactive modes and thus can not be detected by Raman spectrometry. One of the two $A_{2u}$ phonons is an out-of-plane stretching mode including both Cr and I atoms vibration, as shown in Fig. \ref{z-optical modes-b}(b); the other $A_{2u}$ phonon is an in-plane anti-phase breathing mode shown in Fig. \ref{in-plane modes-2b}(a). They should be detected by IR optical spectrometry and would change their frequencies with magnetic order variation, since Wang et.al have found that the frequencies of two IR modes are significantly influenced by the magnetic configuration.\cite{Magnetic-phonon-CrI3-2021} The double-degenerate $E_{u}$ phonons shown in Fig. \ref{z-optical modes-b}(d) are three-phase stretching modes only containing motions of nonmagnetic I atoms and can not couple to the magnet order, whereas the other two $E_{u}$ modes shown in Fig. \ref{in-plane modes-2b}(c) and Fig. \ref{in-plane modes-1b}(d) are breathing and rocking modes respectively, which are all including the motion of magnetic Cr atoms and thus can directly couple to the magnet order of X$_3$-Cr$_2$-X$_3$monolayer. It is pointed out that our conculsions are consistent with the experimental infrared spectra of CrI$_3$ monolayer, where all five IR modes have been observed.\cite{tomarchio2021-IR-exp} 
 
In stark contrast, for Janus Y$_3$-Cr$_2$-X$_3$ ( such as F$_3$-Cr$_2$-I$_3$) monolayer, all IR active modes obtain the R activity. Especially, the out-of-plane stretching $A_{1}$ mode including Cr atoms shown in Fig. \ref{z-optical modes-b}(b) transforms from an IR active $A_{2u}$ mode of CrI$_3$ monolayer to an IR and R double-active $A_{1}$ mode of F$_3$-Cr$_2$-I$_3$ monolayer shown in Fig. \ref{z-optical modes}(b). Its frequency is higher than that of the corresponding $A_{2u}$ mode in CrI$_3$ and lower than that in CrF$_3$ monolayer. Therefore, this out-of-plane stretching $A_{1}$ mode can behave as a characteristic peak in optical spectra of Janus Y$_3$-Cr$_2$-X$_3$ monolayer. Besides, its frequency variation with magnetic order may also be detected by comparison of Raman spectra of FM and AFM Y$_3$-Cr$_2$-X$_3$ monolayer. Contastively, the $E_{u}$ mode in Fig. \ref{z-optical modes-b}(d) transform to $E$ mode shown in Fig. \ref{z-optical modes}(d) is both IR and R active too, owing to the absence of Cr atoms, its frequency is insensitive to the magnetic order of F$_3$-Cr$_2$-I$_3$ monolayer. The left two $E$ modes shown in Fig. \ref{in-plane modes-2}(c) and Fig. \ref{in-plane modes-1}(d) include the motion of Cr atoms and thus couple to the magnetic order, which can be reflected by peak splitting in both infrared and Raman spectra of magnetic Janus Y$_3$-Cr$_2$-X$_3$ monolayer.

\subsection{Magneto-optical effect of R active phonons $A_{1g}$ and $A_{1}$}

The magneto-optical effect can be observed in a non-magnetic system under an external magnetic field or magnetic (such as FM and AFM) system. In principle, time-reversal symmetry breaking can produce an extra antisymmetric off-diagonal component to Raman tensor of $A_{1g}$ and $A_{1}$ shown in Table. \ref{Raman tensor-1}. It is this antisymmetric component that induces the polarization rotation of the scattered light. The magneto-optical Kerr effect of $A_{1g}$ phonons (as shown in Fig. \ref{z-optical modes-b}(c)) due to the FM order has been observed in CrI$_3$.\cite{huang2020tuning-via-symmetry-CrI3} The polarization rotation in this magneto-optical Kerr effect induced by a monolayer FM insulator is extraordinarily large, being two orders of magnitude large than that from the magneto-optical Raman effect induced by an external magnetic field.\cite{huang2020tuning-via-symmetry-CrI3}  

In contrast, the magneto-optical Kerr effect of $A_{g}$ mode in FM CrBr$_3$ monolayer is negligible while the magneto-optical Raman effect is significant and even named as giant magneto-optical Raman effect.\cite{Chial-phonon-CrBr3-2021} This implys that the mechanism of the magneto-optical Kerr effect resulting from FM order may be sharply different from the magneto-optical Raman effect induced by the external magnetic field and is worthy of in-depth exploration. Consider that the $A_{1}$ irreducible representation in $C_{3v}$ has the same Raman tensor as that of that of $A_{1g}$ in $D_{3d}$, one can expect to observe the same magneto-optical effect of the corresponding $A_{1}$ mode shown in Fig. \ref{z-optical modes}(c). The $A_{2u}$ of X$_3$-Cr$_2$-X$_3$ monolayer shown in Fig. \ref{z-optical modes}(b), though with also the out-of-plane vibration, is R inactive and thus has no magneto-optical effect. However, for Janus Y$_3$-Cr$_2$-X$_3$ monolayer without inversion symmetry, it tranforms into $A_{1}$ shwon in Fig. \ref{z-optical modes-b}(b), which is R active and thus can also exhibit magneto-optical effect. Furthermore, it is in sharply contast to $A_{1}$ mode shown in Fig. \ref{z-optical modes-b}(c), because it cantains the vibration of magnetic Cr atoms. Thus one can expect to observe a remarkable magneto-optical effect of this $A_{1}$ mode in Janus Y$_3$-Cr$_2$-X$_3$ monolayer in FM phase. This reinforces our conclusion in above subsection that the $A_{1}$ shwon in Fig. \ref{z-optical modes-b}(b) can behave as a characteristic peak in optical spectra of Janus Y$_3$-Cr$_2$-X$_3$ monolayer.
 
\subsection{Chiral phonons and their effects}
In crystal with $C_3$ symmetry, chiral phonons at high-symmetry points have pseudo angular momentum ($l_{ph}=0,\pm1$) defined by transform property under $\hat{C}_3$ rotational operator(rotation by $2\pi/3$ about $z$ axis): $\hat{C}_3u(\textbf{r})=e^{(-i\frac{2\pi}{3}l_{ph})}u(\textbf{r})=\omega^{(-l_{ph})}u(\textbf{r})$.\cite{Zhang2015Chiral-phonon} The phonon mode with $l_{ph}=0$ has 1D irreducible representation, while the corresponding $l_{ph}=\pm1$ modes have 2D irreducible representation if the inversion symmetry is not broken for non-magnetic system.
 
For X$_3$-Cr$_2$-X$_3$ monolayer, the $E_{g}$ modes shown in Fig. \ref{z-optical modes-b}(e), \ref{in-plane modes-2b}(d), and \ref{in-plane modes-1b}(c) are chiral phonons with $l_{ph}=\pm1$, which chirality is permitted to be observed in helicity-resolved Raman scattering. The $E_{u}$ modes shown in Fig. \ref{z-optical modes-b}(d) and Fig. \ref{in-plane modes-2b}(c) can also be observed directly in infrared absorption optical spectra. In practice, all the $E_{g}$ modes would mix to some extent, and all the $E_{u}$ modes would mix too, which may suppress the chirality of phonon, especially for chiral phonons at $\Gamma$ point. Nevertheless, the chiral phonon and its complete switch of the polarization of an incident circularly polarized light involved in a Raman scattering process has been observed in CrBr$_3$ 2D magnet,\cite{Chial-phonon-CrBr3-2021} which indicates that the incident circularly polarized light may induce or enhance the chirality of doubly degenerate phonons. Therefore we can expect to observe chiral phonons in other 2D X$_3$-Cr$_2$-X$_3$ monolayer.

For Janus Y$_3$-Cr$_2$-X$_3$ monolayer, all the $E$ modes, such as shown in Fig. \ref{z-optical modes}(d,e) and Fig. \ref{in-plane modes-2}(c,d),  are chiral phonons with pseudo angular momentum $l_{ph}=\pm1$, whose chirality should also be observed in both helicity-resolved Raman scattering and infrared absorption optical experiments. These chiral phonons are even easier to be observed than that in the X$_3$-Cr$_2$-X$_3$ monolayer because of the absence of inversion symmetry.
 
In addition, the $E$ modes including Cr atoms in Fig. \ref{in-plane modes-1}(d) and Fig. \ref{in-plane modes-3}, or the $E_{u}$ mode in Fig. \ref{in-plane modes-1b}(d) as well as the $E_{g}$ mode in Fig. \ref{in-plane modes-3b}, are in-plane shear modes, which two partners ($u_{1}, u_{2}$) can be combined into chiral phonons with partners $u_{1}\pm iu_{2}$, which reprsent the circular motion of Cr atoms. These chiral phonons not only have pseudo angular momentum but also have real angular momentum,\cite{zhang2022chiral} while their non-zero real angular monentum only emergent in system without time reversal or space inversion symmetry. The orbit magnetic moment of these chiral phonons should be coupled to the spin magnetic moment of Cr atom in magnetic Y$_3$-Cr$_2$-X$_3$ and X$_3$-Cr$_2$-X$_3$, which would lead to the degeneracy lifting of 2D chiral phonons in FM phase. Furthermore, for the $E_{u}$ mode in Fig. \ref{in-plane modes-2b}(c) with inversion symmetry, its vibration includes the circular motion of Cr atoms, thus the angular momentum can realize only in magnetic f X$_3$-Cr$_2$-X$_3$ monolayer; while for the corresponding $E$ mode in Fig. \ref{in-plane modes-2}(c), its non-zero angular monentum can realize also in mon-magnetic Y$_3$-Cr$_2$-X$_3$ monolayer due to the absence of inversion symmentry. 

We find that there are two distinct types of chiral phonons at $\Gamma$ for Y$_3$-Cr$_2$-X$_3$ monolayer with (X, Y) $\in$ \{F, Cl, Br, I\}, both of them have 2D irreducible representation, one with only pseudo angular momentum but the other also with real angular momentum. This seems inply that the chirality of phonon dose not always mean the circular polarization of phonon, in other word, the circular motion of involving atoms. The chirality would be exhibited in helicity-resolved optical spectra even for non-magnetic system with spatial inversion symmetry,  while the non-zero angular momentum of phonons would lead to spectral splitting of degenrated chiral phonons due to the coupling of its magnetic momentum to the spin of magnetic atoms. Thus we conclude that the chirality of phonon is determined by its pseudo rather than real angular monontum.

\section{Conclusions}
In conclusion, we have performed a group theory investigation on the optical phonons in monolayer X$_3$-Cr$_2$-X$_3$ and their Janus structures Y$_3$-Cr$_2$-X$_3$. The irreducible representation of all phonons at $\Gamma$ point as well as the IR and R activity of optic phonons have been obtained. The Raman tensors and the linear polarization assignments of Raman active phonons as well as the compatibility relation of phonon irreducible representation between Y$_3$-Cr$_2$-X$_3$ and X$_3$-Cr$_2$-X$_3$ monolayer are offered, which provides essential guidance for further optical spectral identification of Janus Y$_3$-Cr$_2$-X$_3$. Furthermore, the corresponding phonon eigenfunctions and their corresponding schematic representations facilitate the experimental identification of their optic phonons. We conclude that: (1) the out-of-plane stretching $A_{1}$ mode including both Cr and halogen atoms may be manifested as a characteristic peak in optical spectra of Janus Y$_3$-Cr$_2$-X$_3$ monolayer; (2) the giant magneto-optical effect of the corresponding $A_{1}$ mode in FM Janus Y$_3$-Cr$_2$-X$_3$ monolayer should be observed in polarized Raman spectra experiment; (3) the chiral phonons at $\Gamma$ in Y$_3$-Cr$_2$-X$_3$ monolayer should be revealed in helicity-resolved optical spectra. Our work enriches the detailed understanding of optic phonons in Y$_3$-Cr$_2$-X$_3$ monolayer with (X, Y) $\in$ \{F, Cl, Br, I\}, and offers theoretical guidance for exploring their distinct physical effects experimentally.

\begin{acknowledgments}
This work was supported by the doctoral research start-up funds of Teacher in Xi'an University of Technology (Grant No.109-451119001) and partly by the Natural Science Basic Research Program of Shaanxi (Program No. 2022JQ-063). H. B. Niu acknowledges the financial support of the Natural Science Foundation of Shaanxi Province, China (No. 2021JM-541). V. Wang acknowledges the financial support of the National Natural Science Foundation of China (Grant No. 62174136) and the Youth Innovation Team of Shaanxi Universities.
\end{acknowledgments}
\nocite{*}
\bibliographystyle{aipnum4-1}
\bibliography{References}
\end{document}